\documentclass{ws-procs975x65}
\newcommand{\ve}[1]{\mbox{\boldmath $#1$}}

\begin{document}

\title{NUMERICAL RELATIVITY SIMULATIONS OF MAGNETIZED BLACK
  HOLE---NEUTRON STAR MERGERS}

\author{ZACHARIAH B. ETIENNE$^*$, YUK TUNG LIU, VASILEIOS PASCHALIDIS,
  and\\ STUART L. SHAPIRO}

\address{Department of Physics, University of Illinois,\\ 
Urbana, IL 61801-3080, USA\\
$^*$E-mail: zetienne@illinois.edu\\
www.illinois.edu}

\begin{abstract}
We present new numerical techniques\cite{etie2011a}
we developed for launching the first parameter study of {\it
  magnetized} black hole--neutron star (BHNS) mergers, varying the magnetic fields seeded
in the initial neutron star. We found that
magnetic fields have a negligible impact on the gravitational
waveforms and bulk dynamics of the system during merger,
regardless of magnetic field strength or BH spin. In a recent
simulation, we seeded the remnant disk from an 
unmagnetized BHNS merger simulation with large-scale, purely poloidal magnetic
fields, which are otherwise absent in the full simulation. The outcome
appears to be a viable sGRB central engine.
\end{abstract}


\bodymatter

\section{Evolving the Equations of General Relativistic
  Magnetohydrodynamics on Adaptively-Refined Grids}

We have recently extended our adaptive mesh refinement (AMR) numerical
relativity code to solve the ideal magnetohydrodynamics (MHD)
equations, enabling us to simulate MHD effects in dynamical spacetimes
with AMR\cite{etie2011a,etie2011b}. The subtlety in evolving these
equations is enforcing the divergence-free constraint $\ve{\nabla}\cdot
\ve{B}=0$. If we were to evolve the induction equation in the most
obvious and straightforward way, numerical errors will lead to
violation of this divergence-free constraint, resulting in the
production of spurious magnetic monopoles. There are
several known solutions to this problem in unigrid simulations, but
few when AMR is used. The one we chose was to evolve the vector
potential, $\mathcal{A}^{\mu}$. In this case, the magnetic induction
and divergence-free equations in curved spacetime become:

\begin{eqnarray}
B^i &=& \epsilon^{ijk} \partial_j A_k, \\
\partial_t A_i &=& \epsilon_{ijk} v^j B^k - \partial_i (\alpha \Phi -
\beta^j A_j) 
\label{A_evol_eqn}
 \\
\partial_j \tilde{B}^j &=& \partial_i( \tilde\epsilon^{ijk} \partial_j A_k)
\equiv 0, \label{divB}
\end{eqnarray}
where $B^i=\tilde B^i/\sqrt{\gamma}$ is the magnetic field measured by a normal observer, 
 $A_{\mu}=\mathcal{A}_{\mu}-\Phi n_{\mu}$ is the projection of
the four-vector potential $\mathcal{A}_{\mu}$ onto a 3-dimensional spacelike
hypersurface, $\Phi$ the scalar potential, $n^{\mu}$ is the normal vector to the hypersurface, 
$\tilde\epsilon^{ijk}=\epsilon^{ijk}/\sqrt{\gamma}$,
$\epsilon^{ijk}=n_{\mu}\epsilon^{\mu ijk}$ is the 3-dimensional
Levi-Civita tensor associated with the 3-metric $\gamma_{ij}$, 
and $\gamma$ the 3-metric determinant.

By construction, we {\it guarantee} in Eq.~\ref{divB} that
the divergence of the magnetic field is zero, since the divergence of
a curl is zero. This property is guaranteed, no matter what
interpolation scheme we choose when interpolating between different
adaptively refined grids.

When evolving the vector potential, an electromagnetic (EM)
gauge choice must be made.  In choosing an EM gauge, there is a
subtlety. We have written our numerical prescription so that the
resulting magnetic fields are completely invariant to the
EM gauge choice inside uniform-resolution grids.

However, when we adaptively add subgrids at higher resolution using
AMR, {\it interpolation at mesh refinement boundaries turns EM gauge
modes into physical modes}, thereby affecting the magnetic fields.
Thus, if we are not careful in our gauge choice, the
gauge-dependent magnetic fields induced on these refinement boundaries
may poison our simulation.

Our first attempt at a gauge condition was $\partial_i (\alpha
\Phi-\beta^j A_j)=0$, as it greatly simplifies the right-hand side
of Eq.~\ref{A_evol_eqn}. However, we later found that this gauge
choice introduces a zero-speed gauge mode\cite{etie2011a}. With this
zero-speed mode, if the path of magnetized matter crosses an AMR
refinement boundary, interpolation on this boundary leads to the
creation of weak, spurious magnetic fields in black hole--neutron star
(BHNS) simulations that grow stronger with time until the simulation
crashes.

So we switched from our original choice to the Lorenz gauge
$\nabla_{\mu} \mathcal{A}^{\mu}=0$, in which the EM gauge modes
propagate away, thereby drastically reducing the appearance of
spurious magnetic fields at refinement boundaries. The simulations
presented in the next section were the first to use this gauge for
full GRMHD with AMR.

\section{Magnetized Black Hole---Neutron Star Binary Mergers}

As a neutron star (NS) is tidally disrupted by a black hole (BH)
companion at the end of a BHNS binary inspiral, its magnetic fields
will be stretched, wound, and amplified. If sufficiently strong, these
magnetic fields may impact the gravitational waveforms, merger
evolution and mass of the remnant disk. Formation of highly-collimated
magnetic field lines in the disk+spinning BH remnant may launch
relativistic jets, providing the central engine for a short-hard GRB (sGRB). We
explore this scenario through fully general relativistic,
magnetohydrodynamic (GRMHD) BHNS simulations from inspiral through
merger and disk formation\cite{etie2011b}. In particular, we attempt
to answer the following two questions:
\begin{enumerate}
\item How do NS magnetic fields affect BHNS binary waveforms and the resulting BH+disk system?
\item Do we produce an sGRB progenitor?
\end{enumerate}

\begin{figure*}
\epsfxsize=2.4in
\leavevmode
\epsffile{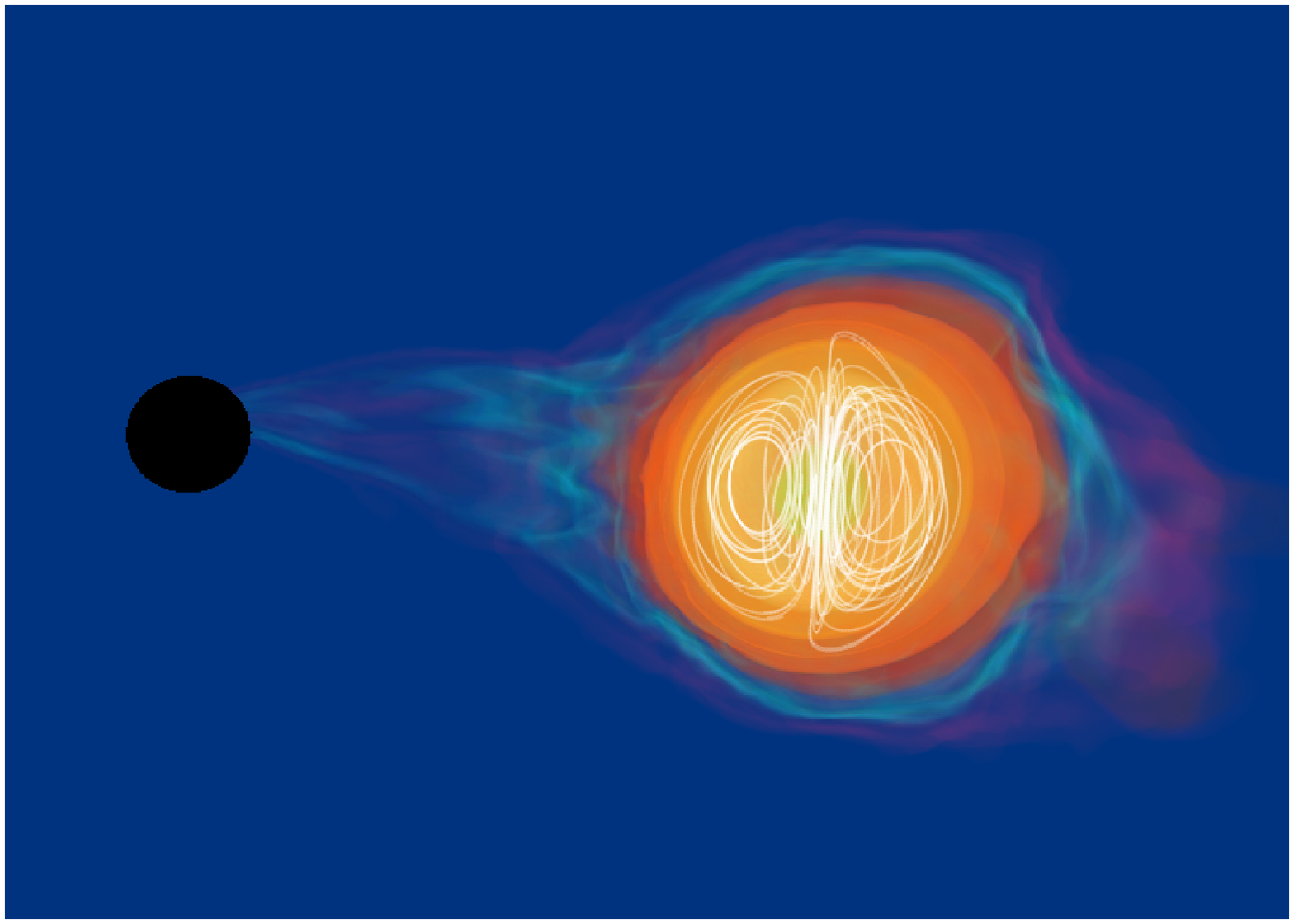}
\epsfxsize=2.4in
\leavevmode
\epsffile{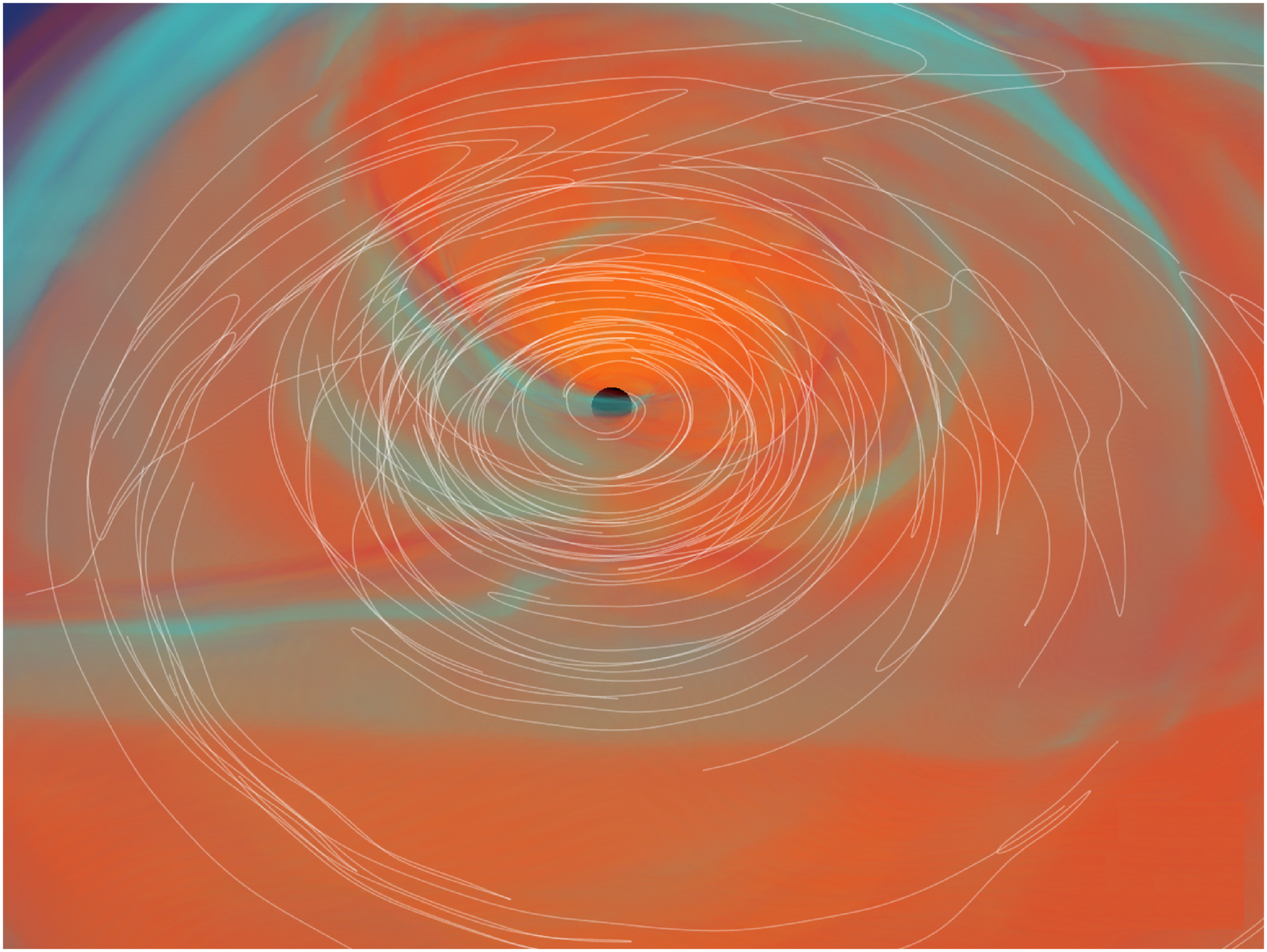}
\caption{3D density and magnetic field snapshots. Left: initial data, NS on
  the right (from highest to lowest rest-mass density, the colors are:
  yellow, orange, red, and cyan), BH apparent horizon (AH) on the
  left. Right: final disk density profile with magnetic field
  lines, about $33$ms $(1.4 M_{\odot}/M_0)$ after disk
  formation ($t=2072M$), where $M_0$ is the initial rest mass of the NS and $M$
the ADM mass of the system.
}
\label{figbasicstory}
\end{figure*}

To answer these questions, we perform simulations in
which the BH is initially spinning with spin parameter 0.75, aligned
with the orbital angular momentum. Though {\it surface} NS magnetic field
strengths have been inferred by observation, very little is known
about NS {\it interior} magnetic fields. So for this preliminary
investigation we seed only the NS interior with initially poloidal
magnetic fields. The initial data are shown in the left panel of
Fig.~\ref{figbasicstory}. Keeping this field configuration fixed, we
vary the initial magnetic field strength, choosing magnetic fields
with average magnetic to gas pressure $P_{\rm B}/P_{\rm gas}$ of 0,
$5\times10^{-5}$, and $5\times10^{-3}$. Note that the case with the
strongest magnetic fields has field strengths of order $10^{17}$G at
the core of the NS (assuming the NS has a rest mass of
1.4$M_{\odot}$). We choose to seed the NS with 
magnetic fields sufficiently weak to avoid disturbing the NS equilibrium during
inspiral, but sufficiently strong to influence the final outcome.

To address the first question, we find that magnetic fields have no
significant impact on the gravitational waveforms or
residual disk masses, regardless of initial strength. 
Magnetic fields retain their poloidal structure
during the final orbit before merger. But in terms of magnetic field
structure, there is a large difference between pre- and post-disruption: 
magnetic fields that were almost purely poloidal initially
become almost purely toroidal due to the rapid winding of
the matter around the BH as it forms a disk.
One of the ingredients in sGRB models is the collimation
of magnetic fields perpendicular to the disk. The right frame of
Fig.~\ref{figbasicstory} demonstrates the lack of magnetic
field collimation in the $P_{\rm B}/P_{\rm gas}=5\times10^{-3}$ case.

We stopped our simulation about 30ms after tidal
disruption. Notice that we had a thick disk orbiting a
spinning BH, but there was no strong evidence of
magnetic field collimation. But what about relativistic outflows,
another key ingredient for sGRB central engines? After 30 ms of disk evolution, 
we find no outflows: low-density matter is still rapidly accreting onto the BH
poles and no sign of fluid velocity reversal is observed.

\begin{figure*}
\begin{center}
\vspace{-4mm}
\epsfxsize=3.2in
\leavevmode
\epsffile{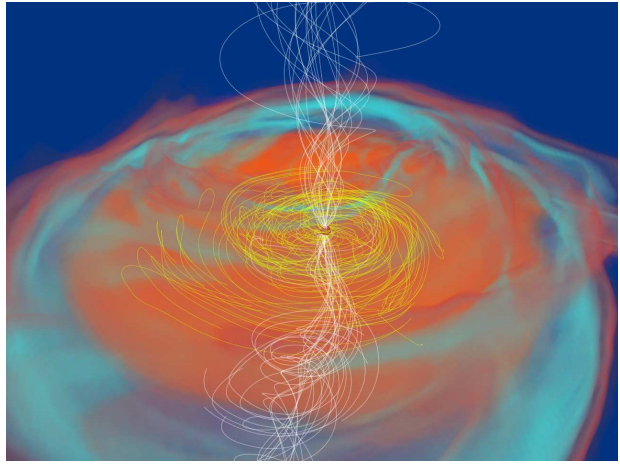}
\caption{3D snapshot, corresponding to the case in which we seed a
  remnant disk from an unmagnetized BHNS simulation with purely
  poloidal magnetic fields. This is a snapshot taken when we terminate the
  simulation, viewing from above the disk
  plane. Magnetic field streamlines emerging just above
  and below the BH poles are shown in white, and those in the disk are shown
  in yellow.}
\label{meridionaldiskB5}
\end{center}
\end{figure*}

In our latest
work\cite{etie2012}, we demonstrated that when the
remnant disk from an unmagnetized BHNS simulation is seeded with
large-scale poloidal fields, we observe spectacular collimated
magnetic fields and relativistic outflows, as shown in the panels of
Fig.~\ref{meridionaldiskB5}. However, such large-scale poloidal fields may be difficult
to generate in a fully self-consistent BHNS simulation, as the
magnetic fields must follow the NS fluid as it wraps around the spinning
BH during tidal disruption and disk formation, generating strong
toroidal fields. GRMHD simulations performed by other groups indicate that BH
accretion disks lacking large-scale poloidal fields may not be capable
of generating sustained jets\cite{GRMHD_Jets_Req_Strong_Pol_fields}.
This result combined with our findings, make BHNS mergers less likely
sGRB central engines. 

In spite of this, we found in this same work\cite{etie2012} that 
inserting {\it tilted} magnetic fields into the NS breaks the
initial equatorial symmetry of the problem and
encourages poloidal fluid motion, resulting in 10x stronger poloidal
magnetic fields in the remnant disk. Even with these stronger poloidal
magnetic fields, no magnetic collimation or relativistic outflows were
observed. We anticipate that large-scale poloidal fields might be
produced in BHNS simulations with highly-spinning BHs that are
misaligned with the orbital angular momentum. Such a system might
explain quasi-periodic signals in observed sGRBs\cite{stone2012}.

\bibliographystyle{ws-procs975x65}
\bibliography{main}

\end{document}